\documentclass[twocolumn]{aastex7}
\usepackage{hyperref}
\usepackage{amsmath}
\newcommand{\fracbrac}[2]{\left(\frac{#1}{#2}\right)}
\newcommand{\paren}[1]{\left( #1 \right)}

\usepackage[utf8]{inputenc}
\usepackage{float}
\usepackage{natbib,ulem,subcaption}

\begin{document}
\title{
Free Floating or Merely Detached?
}
\date{July 2025}
\author[0000-0002-1032-0783]{Sam Hadden}
\affiliation{Canadian Institute for Theoretical Astrophysics, 60 St George St Toronto, ON M5S 3H8, Canada}
\email{hadden@cita.utoronto.ca}
\author[0000-0003-0511-0893]{Yanqin Wu}
\affiliation{Department of Astronomy \& Astrophysics, University of 
Toronto, Toronto, Canada}
\email{wu@astro.utoronto.ca}

\begin{abstract}
Microlensing surveys suggest the presence of a surprisingly large population of free-floating planets, with a rate of about two Neptunes per star. The origin of such objects is not known, neither do we know if they are truly unbound or are merely orbiting at large separations from their host stars. Here, we investigate planet-planet scattering as a possible origin through  numerical simulations of unstable multi-planet systems. We find that planet ejection by scattering can be slow,  often taking more than billions of years for Neptune-mass scatterers orbiting at a few AU and beyond. Moreover, this process invariably delivers planets to orbits of hundreds of AU that are protected from further scattering. We call these  ``detached" planets. Under the scattering hypothesis, we estimate that about half of the reported ``free-floating" Neptunes are not free but merely ``detached".
\end{abstract}

\section{Introduction}
\label{sec:intro}
Gravitational microlensing surveys are uniquely capable of discovering planetary mass objects at large distances from their host stars \citep{Mao1991}. This includes the so-called ``free-floating planets", where no lensing signature of the host stars are detected \citep[FFPs; e.g.,][]{sumi_unbound_2011,mroz_no_2017,mroz_two_2019,mroz_terrestrial-mass_2020,kim_kmt-2019-blg-2073_2021,ryu_kmt-2017-blg-2820_2021,gould_free-floating_2022,koshimoto_terrestrial-_2023}.  Whether they are truly `free' or just orbiting at very large distances is not known. For a recent review, see \citet[][]{mroz_exoplanet_2023}.

Recently, based on about a dozen short-timescale microlensing events  collected over 9 years of  the MOA-II microlensing survey, \citet{sumi_free-floating_2023} deduced the following occurrence rate for FFPs,  
\begin{equation}
    {{dN}\over{d\log_{10} M_p}} =
{2.18^{+0.52}_{-1.40}}
\times \left({m_p\over{8 M_\oplus}}\right)^{-p} \, ;\hskip0.1in p = 0.96^{+0.47}_{-0.27}\, .
\label{eq:sumi}
    \end{equation}
This is the rate per star and applies for planet masses ($m_p$) between  $0.3 M_\oplus$ and $20M_J$. This occurrence rate implies a surprisingly large reservoir of FFPs: integrating over the above mass range, one obtains $20^{+23}_{-13}$ FFPs per star, corresponding to a total mass of $\sim 80~ M_\oplus$.
Though surprising, these results corroborate earlier results by \citet{mroz_no_2017} and \citet{gould_free-floating_2022}. Here, we will focus on producing free-floating `Neptunes', planets with planet-to-star mass ratios similar to that of Neptune to our Sun, or  $M_p \approx 8 M_\oplus$ for a typical M-dwarf of $M_* = 0.5 M_\odot$.  Integrating over the mass decade around this value, the lensing result can be cast  more memorably as ``two free-floating Neptunes per star".\footnote{The earlier claim of `two free-floating Jovians per star' \citep{sumi_unbound_2011} has been revised \citep{sikora_updated_2023}.} 

Such a numerous and massive population is even more striking when one considers the fact that a typical galactic star is an M-dwarf. For context, we review the known bound planet population around M-dwarfs:
\begin{itemize}
    \item Planets inward of $1$~AU are most easily characterized by transit. \citet{hsu2020} found an occurrence rate of a couple super-Earth/mini-Neptunes per M-dwarf.
    This corresponds to a solid mass of $\sim 10 M_\oplus$ per star \citep{Mulders2015}. 
    \item Planets between $\sim1-10$AU  are probed by microlensing and radial velocity studies. Based on 63 microlensing events from KMTNet and OGLE, \citet{zang_microlensing_2025} reported a single power-law fit for the mass-ratio distribution of
    \begin{equation}
        {{dN}\over{d\log q}} = (0.18\pm 0.03)\times \, \left({q\over{10^{-4}}}\right)^{-0.55\pm 0.05}\, ,
        \label{eq:zang}
    \end{equation}
    where $q = M_p/M_*$. They further reported evidences for a bi-modal distribution. For the low-mass peak (lying around $q=8 q_\oplus$), they found $\sim 0.35$ planets per star, or a solid mass of $\sim 1 M_\oplus$ per star. The higher-mass peak, interpreted as giant planets, is about ten times lower  in number, consistent with the RV-reported occurrence of giant planets around M-dwarfs \citep[$6.5\%\pm 3.0\%$,][]{montet_trends_2014}.

\item  While we have no ready probe for planets lying beyond $10$~AU, one is made to believe that nascent planetary systems do not stretch much beyond $\sim 30$~AU, the characteristic sizes for proto-planetary disks \citep[e.g.][]{Tobin2020}.

\end{itemize}

At face value, the FFPs as reported by \citet{sumi_free-floating_2023} are much more numerous and contain much more mass than the bound planet population. They are also, on average, lighter.

Are the FFPs formed in isolation, or are they formed in  protoplanetary disks? Isolated formation may account for a portion of FFPs with masses in the regime of giant planets and brown dwarfs. Direct gravitational collapse is thought to be capable of producing objects as small as $\sim 0.005 M_\odot$ \citep{Whitworth2004}, while disruption of accretion by dynamical interactions \citep{Reipurth2001} or photo-erosion \citep{Whitworth2004} may enable the formation of somewhat  lighter objects. Lower mass FFPs, by contrast, are thought to form exclusively in protoplanetary disks. 

 If these objects are formed in disks, why are their host stars' lensing signatures missing?  It could be that these planets are located outside a few stellar Einstein radii from the stars and are therefore lensing independently. This would require orbital separations beyond $\sim 10$AU. But squeezing many planets between this radius and the outer edge of the proto-planetary disks ($\sim 40$AU)--- ten times more than those found in the 0.1-1~AU and 1-10~AU ranges--- seems dynamically too crowded (also see  Section \ref{sec:discussion}). It seems more likely that most of the FFPs are the results of some dynamical ejection mechanism. Potential mechanisms include interactions with stellar binary companions \citep[e.g.,][]{Kaib2013},
stellar flybys in dense stellar environments
    \citep{
        laughlin_modification_1998,
        bonnell_planetary_2001,
        Malmberg2011,
        parker_effects_2012,
        cai_stability_2017,
        van_elteren_survivability_2019,
        rodet_correlation_2021,
        yu_free-floating_2024},
host star mass loss during post main-sequence evolution \citep{Veras2011},
and planet-planet scattering following orbital instability \citep[e.g.,][]{rasio_dynamical_1996,weidenschilling_gravitational_1996,lin_origin_1997,juric_dynamical_2008,veras_planet-planet_2012,Gautham2025}.  

Among these, we choose to focus on planet-planet scattering. Typical flybys in star clusters reach close-approaches of order $10^2-10^3$ AU. This is too far to directly eject planets that orbit at tens of AU, though it could destabilize them and promote planet-planet  scattering \citep{Malmberg2011,boley_interactions_2012}. Binary ejection, on the other hand, is subject to much uncertainty. It is unclear how planets form around binaries, though evidence suggests that planet formation is suppressed around binaries with separations between a few to a few tens of AU \citep{Moe2021}. Binaries that are much closer or much wider may behave effectively like single stars \citep[but see][]{coleman_properties_2024,Coleman2025}.

In light of the rarity of giant planets,  we  extend previous scattering studies that mostly focus on giant planets (see above) down to lower masses.  We simulate planet-planet scattering in systems comprised of equal-mass planets across a range of mass scales.  We also explore how a single Jupiter interacts with many Neptunes. We find that the timescale over which a scattering system ejects planets depends sensitively on the mass of the  ejector, with lower-mass  ejectors taking much longer. Ejection takes multiple Gyrs for a Neptune mass planet at a distance  of $\sim 10~\mathrm{AU}$.  Moreover, we  point out that such a process usually leaves some planets stranded at large distances -- we call these the `detached' planets. This leads us to suggest that a significant fraction of the reported FFPs are likely to be  detached instead.

\begin{figure}
    \centering
    \includegraphics[width=0.9\columnwidth]{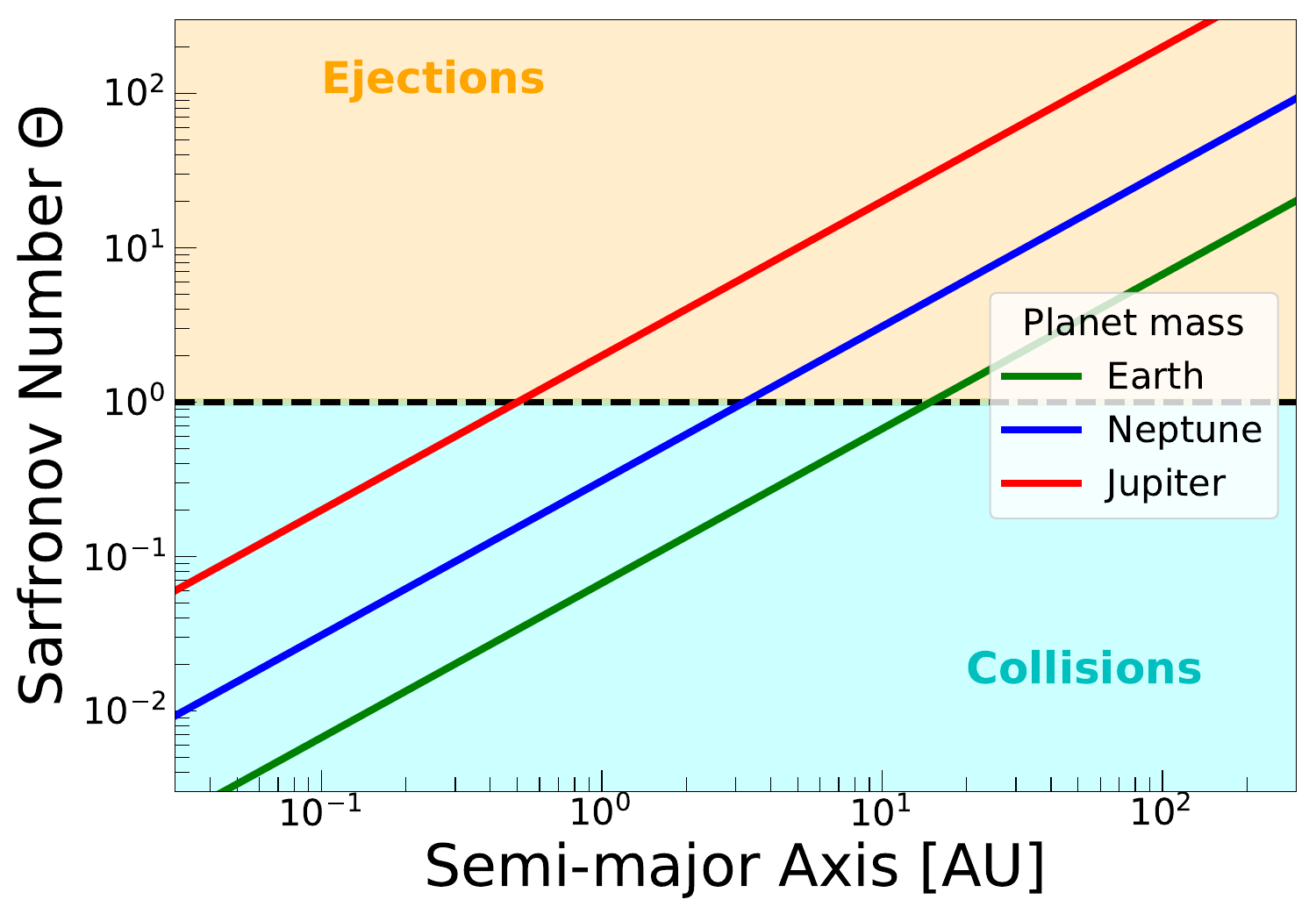}
    \caption{Safronov number, $\Theta$, as a function of semi-major axis for Earth-, Neptune-, and Jupiter-mass planets around a solar mass star. Close encounters among planets with Safronov numbers less than unity will generally lead to  collisions, while to ejections in the opposite limit.}
    \label{fig:safronov}
\end{figure}

The  key quantity that determines the ultimate fate of scattering systems is the `Safronov number' \citep{Safronov1972}. It is 
\begin{equation}\label{eq:safronov}
    \Theta = \frac{v_{p,\mathrm{esc}}^2}{2V_\mathrm{orb}^2} = 
    \frac{m_p}{R_p}\frac{a_p}{M_*}~,
\end{equation}
where $v_{p,\mathrm{esc}}$ is the escape speed form a planet's surface, $V_\mathrm{orb}$ is the velocity of a circular orbit at the planet's orbital location, $R_p$ is the planet's radius, $a_p$ is its semi-major axis, and $m_p$ and $M_*$ are the planet's and star's mass, respectively. When $\Theta \gg 1$, planet encounters can easily eject bodies from the star's potential well. While when $\Theta \ll 1$, collisions rather than ejections dominate.
Figure \ref{fig:safronov} illustrates the Safronov numbers of an Earth-, Neptune-, and Jupiter-mass planet at different orbital separations from a solar-mass star. Neptune-mass planets can eject outwards of a few AU, while Earth-like planets almost always end up in collisions. 

\section{Numerical Integrations}
\label{sec:numerical}
We take the mass of the host star to be $1M_\odot$ in our simulations.  It is  straightforward to scale for a different value. We treat all bodies in our simulations as point masses with no physical extent, ignoring  possible  collisions between planets as well as between planets and the star.  This makes our simulations scale-free, though our results only apply in the $\Theta \geq 1$ regime. We report all distances normalized by the initial semi-major axis of the innermost planet, $a_{1,0}$, along with timescales normalized  by its initial orbital period, $P_{1,0}$. 

We  first integrate planetary systems of five equal-mass planets.  This includes four ensembles, comprised of 101 numerical integrations each, with planet masses set to $1, 3, 10$ and $30$  Neptune masses ($m_\mathrm{Nep}$). Each integration lasts for $10^8$ orbits. Planets are initialized on circular orbits with orbital inclinations drawn randomly from a Rayleigh distribution with a mode of $10^{-3}$ radians.
Mean longitudes and longitudes of ascending node are drawn randomly from a uniform distribution between $0$ and $2\pi$.   Adjacent planets have constant fractional spacing in semi-major axis. For the two lowest mass ensembles, planet separations were chosen to produce orbital instabilities on timescales distributed uniformly in logarithm between $10^4$ and $10^5~P_{1,0}$, according to Equation 6 of \citet{Lammers2024}. This yields fractional orbital separations on the order of $\Delta a/a \sim 0.1$. For the two higher-mass ensembles, the targeted range of instability time scales  was adjusted to all between $10^3$ and $10^4$ orbits, corresponding to  a fractional orbital separations of $0.15 \lesssim \Delta a/a \lesssim 0.25$. This adjustment was necessary as initial experiments resulted in a large number of systems failing to destabilize.
We find that some systems fail to destabilize over the course of our $10^8$ orbit integrations. These simulations are discarded from our analysis below, leaving 89 systems in the $3~m_\mathrm{Nep}$ ensemble and 80 systems in the $30~m_\mathrm{Nep}$ ensemble. (All 101 simulations in the $m_\mathrm{Nep}$ and $10~m_\mathrm{Nep}$ ensembles destabilize.) 

To investigate the possible effects of a Jovian planet, we  also integrate an  ensemble of planetary systems comprised of 10 Neptune-mass planets and a single Jupiter (called `1J+10N' below). The systems are initialized with the Jupiter orbiting interior to  all the Neptunes. All initial conditions are set in the same manner as above, with fractional separations between the Jupiter and the first Neptune chosen to be a factor of $\left(\frac{m_\mathrm{Jup}}{m_\mathrm{Nep}}\right)^{1/4}\approx 2.1$ times the spacing  between two Neptune planets.

All numerical integrations are carried out using the \texttt{ias15} integrator \citep{rein_ias15_2015} implemented in the  \texttt{rebound} package \citep{rein_rebound_2012}.  We use \texttt{rebound} version 4.3.1, which implements the criterion described in \citet{Pham2024} for adapting the time step of the \texttt{ias15} integrator.

We monitor the energy error in each simulation.
{75\%} of our simulations of equal-mass planets conserve energy to better than $1\%$. Energy errors larger than this threshold generally arise in simulations in which an exceptionally close encounter between the star and a planet occurs away from the coordinate system origin, which is placed at the system's barycenter. Conservation of momentum in systems that eject planets causes the barycenter of the star and the remaining bound planets to drift away from the coordinate system origin.  Since particle positions are recorded as floating point numbers with a fixed number of digits, there are an insufficient number of digits to accurately resolve these very close planet-star encounters. Fortunately, this is not a serious concern: such extremely close encounters would almost certainly result in planet-star collisions or strong tidal dissipation in realistic planetary systems, removing energy from the system. Below, we report results for both the full ensembles of integrations and for the subsets that conserve energy to better than $1\%$. We find that this energy cut makes little difference in the statistics of the outcomes. Fractional energy errors in all (but one\footnote{The one exception  has a fractional error of $\sim 4\times 10^{-4}$, equivalent to approximately $2\%$ of the initial binding energy of the most strongly-bound Neptune-mass planet. So this is not of serious concern.}) of our 1J+10N simulations  never exceed $10^{-6}$. All fractional errors in angular momentum are  better than $1\%$, and, in 95\% of simulations, they are better than $0.01\%$.
\section{Results}
\label{sec:results}
Here we present the results of our numerical integrations, first for the equal-mass ensembles and then for the 1J+10N ensemble.

\subsection{Equal-mass planets}
\label{sec:equal_mass}

\begin{figure}
    \centering
    \includegraphics[width=0.95\columnwidth]{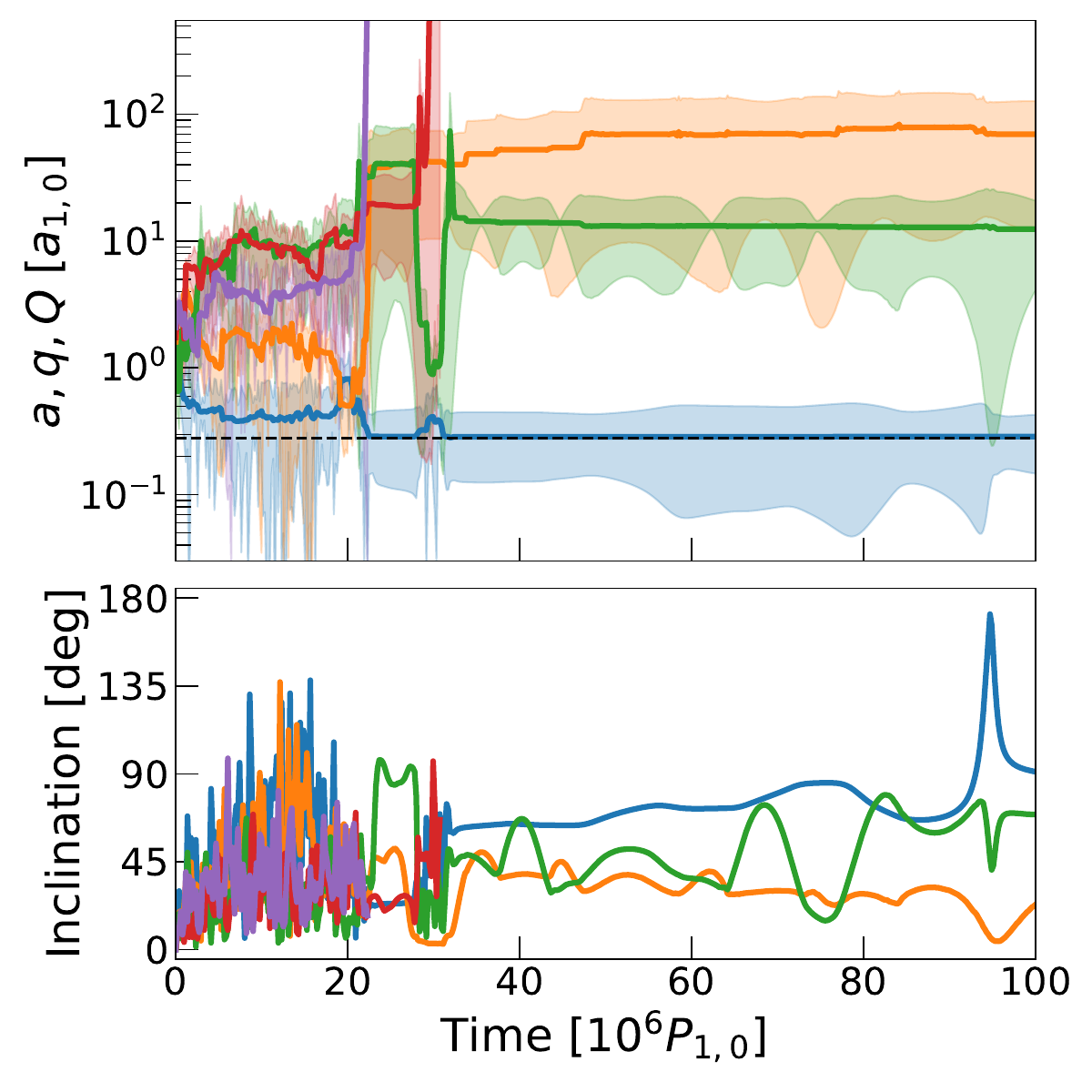}
    \caption{Example evolution of an unstable system with 5 one-Neptune-mass planets. 
    \emph{Top:} the semi-major axes of individual planets are plotted as solid lines. Orbits' radial extents between pericenter and apocenter are indicated by the corresponding shaded regions. The black dashed line indicates the semi-major axis of  an orbit with the same binding energy as the initial five-planet system. 
    \emph{Bottom:} planets' orbital inclinations, measured relative to the system's initial invariable plane, as functions of time.
    }
    \label{fig:example}
\end{figure}

Figure \ref{fig:example} shows a typical example of the dynamical evolution in our simulations. Here, a system of 5 Neptune-mass planets becomes unstable, quickly undergoes orbit-crossing, and enters an extended scattering phase. 
Ejections, however, occur only after  $\sim 10^7$ orbital periods.  
One planet is scattered inward, absorbing most of the system's initial binding energy, while the remaining bound planets are scattered out to distances of $\sim 10$ to $100$ times their initial orbits.
These detached planets continue interacting among themselves. This has two consequences. First, they often lift each other's periapses through mutual torques
and protect each other against immediate scatterings with the inner-most body.
 Second, long-term secular exchanges of orbital angular momentum among them will again drive some of them to high eccentricities. These unlucky planets may encounter the inner planet and  risk being ejected. We observe that this latter process is slow: there are still two planets left at large distances (`detached') at the end of our integration ($10^8 P_{1,0}$).

\begin{figure}
    \centering
    \includegraphics[width=0.9\columnwidth]{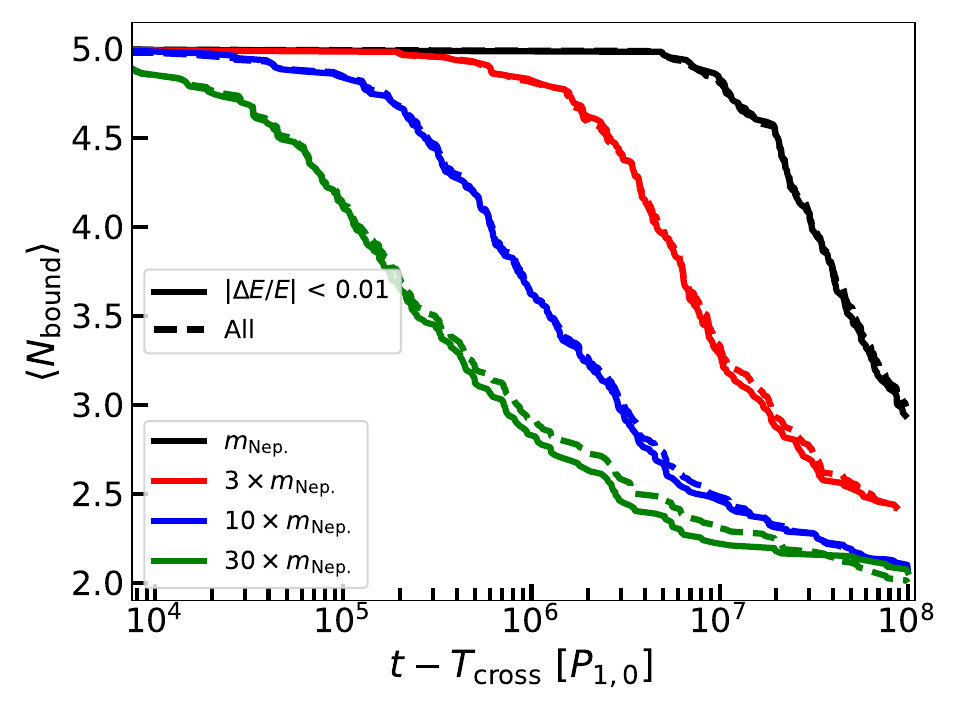}
    \caption{Average number of bound planets as a function of time after first orbit-crossing, plotted in different colors for different ensembles.  Results for all simulations and only those with energy errors $|\Delta E/E|<0.01$ are shown with dashed and solid lines, respectively. Lower-mass ensembles take longer to eject planets. In all cases, more than one planet remains bound.
    }
    \label{fig:bound_vs_t}
\end{figure}

Comparing across different ensembles, we can draw some conclusions on the ejection process.
Figure \ref{fig:bound_vs_t} shows, for each ensemble, the average number of bound planets per system  versus time after the first orbit-crossing.\footnote{We determine the time of the first orbit crossing by evaluating the linking coefficient of \citet{kholshevnikov_linking_1999} for each pair of orbits. This is implemented in  the package \texttt{celmech} {\citep{hadden_celmech_2022}} as \texttt{celmech.miscellaneous.linking\_l}.} 
It is clear that the timescale for planet ejections is strongly mass-dependent, with more massive systems ejecting planets on shorter time scales.
 We fit a power-law to $T_\mathrm{ej}(m_p)$, the time required for the mean number of bound planets to decrease by one and find
\begin{equation}
    T_\mathrm{ej}(m_p) \approx 2.9\times 10^{7}\fracbrac{m_p}{m_\mathrm{Nep}}^{-1.64} P_{1,0}~. \label{eq:Teject}
    \end{equation}
This mass dependence is comparable to but less steep than the decay behaviour of a population of test particles scattered by a single massive perturber. \citet{malyshkin_keplerian_1999,hadden_scattered_2024} showed that, in the latter case, the diffusive evolution of test particles' orbital energies admits a Fokker-Planck description, and the number of bound particles decays over a timescale that is inversely proportional to the square of the planet mass.  In addition to a difference in the mass scaling, Figure \ref{fig:bound_vs_t} shows that the decay in the mean number of planets is algebraic in time, in contrast to the exponential decay seen in test particle scattering. This difference likely arises because, with multiple equal-mass planets, encounters between any two planets can lead to ejection, or the ejection probability increases with the number of bound planets \citep[see, e.g., ][]{adams_migration_2003}.

To further analyze the dynamical evolution in our simulations, we introduce the concept of angular momentum deficit \citep[AMD; e.g.,][]{Laskar1997}.  The total angular momentum deficit of a system hosting $N_\mathrm{bound}$ planets is defined as the negative of the difference between the system's total angular momentum and the angular momentum of a hypothetical system in which the planets are on circular and coplanar orbits with the same semi-major axes. This is
\begin{equation}
    C_\mathrm{tot} = \sum_{i=1}^{N_\mathrm{bound}}m_i\sqrt{GM_*a_i}\left(1-\sqrt{1-e_i^2}\cos I_i\right)~\, ,
\end{equation}
where $m_i, a_i, e_i$ and $I_i$ are, respectively, mass, semi-major axis, eccentricity, and inclination of the $i$th planet, with the inclination measured with respect to the system's invariable plane.
This quantity provides a measure of the overall orbital excitation of a system: it is equal to zero  when all planets' orbits are circular and coplanar and increases monotonically with planets' eccentricities and inclinations (for fixed $a_i$'s).  Interactions that transfer orbital energy between planets (close encounters and resonances) can modify the AMD while under purely secular dynamics it is conserved.

\begin{figure}
    \centering
    \includegraphics[width=\linewidth]{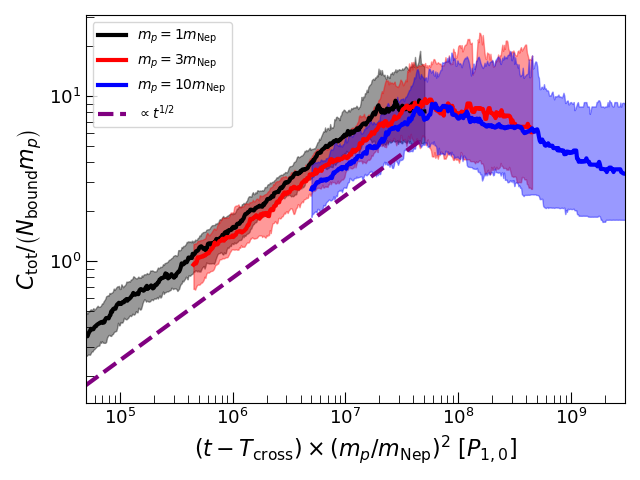}
    \caption{{Systems' AMD per unit mass plotted against time, again in different colors for different ensembles (the $30~m_{\rm Nep}$ ensemble is omitted for legibility). Time is measured from the first orbit-crossing and is scaled by $(m_p/m_\mathrm{Nep})^2$. The solid lines stand for the median value, and the shaded regions show the middle 50\% quantile range.} 
    The dashed purple line indicates a diffusive growth rate ($\propto t^{1/2}$). 
    }
    \label{fig:amd_vs_time}
\end{figure}

Figure \ref{fig:amd_vs_time} shows the evolution of systems' AMD per unit mass, 
\begin{equation}
    \langle\mathrm{AMD}\rangle \equiv
\frac{C_\mathrm{tot}}{N_\mathrm{bound} m_p},
\end{equation} 
in different simulation ensembles. After orbit-crossing, systems' AMDs exhibit diffusive growth so that $\langle\mathrm{AMD}\rangle\propto m_p^2\sqrt{t}$.\footnote{It is straightforward to show from the equations of motion that, in a system of equal-mass planets, $\left\langle\fracbrac{dC_\mathrm{tot}}{dt}^2\right\rangle \propto m_p^4$ and so the AMD per unit mass will exhibit a diffusion rate $\propto m_p^2$.} For all ensembles, this growth continues until the AMD per unit mass reaches a value of $\langle\mathrm{AMD}\rangle \sim 10$, when scattered planets reach distances of $\sim 100 a_{1,0}$.
Beyond this point, a cooling process dominates. Detached planets can gravitationally torque each other and inject AMD into the orbit of some unlucky ones. These latter planets become orbit crossing with the inner-most planet and can be ejected. This tends to remove AMD from the bound planets, dynamically ``cooling" the system.
We turn now to describing how the AMD cooling phase proceeds.

To make this discussion more quantitative, we further introduce the concept of AMD stability, as described by \citet{laskar_amd-stability_2017}.
For the $i$th detached planet in a system, there is a critical  amount of AMD, $C^*_i$,  above which that planet can cross orbits with the inner planet, assuming its semi-major axis is fixed.\footnote{An analytic expression for $C^*_i$, in the special case of equal-mass planets, can be found in Table 1 of \citet{laskar_amd-stability_2017}. 
To compute $C^*_i$ for detached planets in 1J+10N simulations, we use the \texttt{celmech} package's \citep{hadden_celmech_2022} function 
\texttt{miscellaneous.critical\_relative\_AMD}.
} 
The last assumption is reasonable because mutual interactions among detached planets, especially at late times, are dominated by secular interactions. 

If the total AMD ($C_{\rm tot}$) is below $C_*^i$, secular dynamics cannot bring  the planet  into close contact with the inner planet. 
In this case, there are only slow changes in the planet's orbital energy via interactions with other detached planets and we refer to it as AMD-stable.
Conversely,  a larger $C_{\rm tot}$ would potentially allow the planet to  evolve onto a (near-)crossing orbit with the inner planet. This  planet will undergo strong scattering and, eventually, be ejected. 

\begin{figure}
    \centering
    \includegraphics[width=\linewidth]{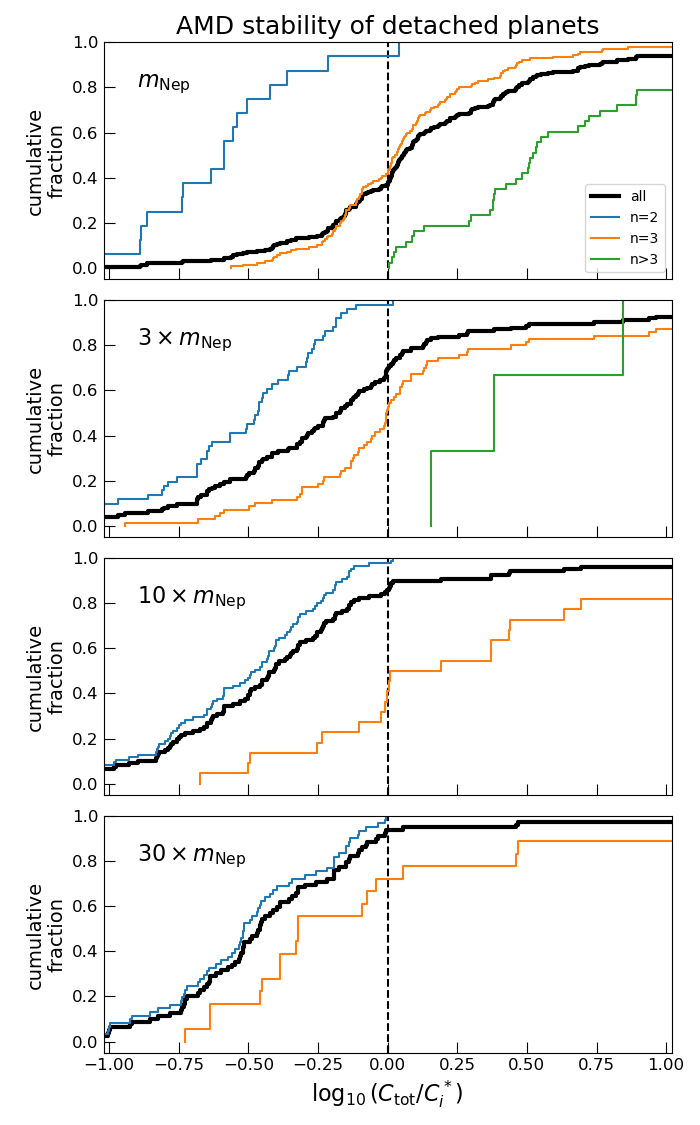}
    \caption{
    Cumulative distributions of the ratios between
    systems' total AMDs, $C_\mathrm{tot}$, and 
    their constituent detached planets' critical AMDs, $C_i^*$, the amount needed to cross orbit with the inner planet. 
    Thick black lines show the cumulative distributions 
    of $C_\mathrm{tot}/C_i^*$ for all detached planets, while colored lines show
    cumulative distributions in systems with $n=2$ (blue), 3 (yellow), or $>3$ (green) planets remaining.
    Detached planets with $C_\mathrm{tot}/C_i^*<1$ are AMD-stable.
    }
    \label{fig:amd_cdfs}
\end{figure}

Figure \ref{fig:amd_cdfs} plots the cumulative distributions of the ratios $C_\mathrm{tot}/ C_*^i$ for detached planets remaining at the end of our simulations. The majority ($82\%$)\footnote{ This fraction rises to $86\%$ if only systems with $|\Delta E/E|<0.01$ are considered.} of systems in the $10~m_\mathrm{Nep}$ and $30~m_\mathrm{Nep}$ ensembles have evolved to a hierarchical configuration of two planets. Figure \ref{fig:amd_cdfs} shows that the lone remaining detached planets in these systems are essentially always AMD-stable. In comparison, only a fraction of the lower-mass ensembles ($m_p = m_\mathrm{Nep}$ and $3~m_\mathrm{Nep}$) have reached this final state and exhibit AMD stability. This is  a consequence of the slower evolution in lower-mass systems.

In those systems with two detached planets (i.e., three planets in total), typically one of them is AMD-stable while the other is marginally AMD unstable with $1<C_\mathrm{tot}/ C_*^i\lesssim 3$.   Given sufficient time, continued exchange of AMD among the planets should eventually cause the AMD-unstable planet to be ejected, carrying away AMD from the system and leaving the other detached planet behind on an AMD-stable orbit. 
For the two lowest-mass ensembles, the fraction of systems still in such an incomplete stage by the end of $10^8$ orbits (or 3~Gyrs for $a_{1,0}=10$AU) is $\sim 60\%$. So we expect a substantial fraction of observed systems to still harbour two detached planets.
A much smaller fraction of systems ($\sim  8\%$ among the lowest-mass ensembles) are still left with three or more detached planets. All of these are AMD unstable (Fig. \ref{fig:amd_cdfs}). They are less further along in their process of AMD cooling and longer integrations should find them ejecting all but one detached planet.

These numerical results suggest that unstable systems of similar-mass planets  evolve until they reach such a hierarchical two-planet configuration, in the ejection regime (high Safranov number) and in the absence of other astrophysical effects.\footnote{They can do so because AMD can be freely exchanged among the planets, likely as a result of strong secular chaos \citep{lithwick_theory_2011}.} This conjecture is supported by previous scattering experiments of \citet[][]{juric_dynamical_2008}, where they reported that systems initially containing ten or 50 giant planets are typically reduced to $\sim 2$ survivors (see their Figure 1).%

\begin{figure}
    \centering
    \includegraphics[width=0.9\columnwidth]{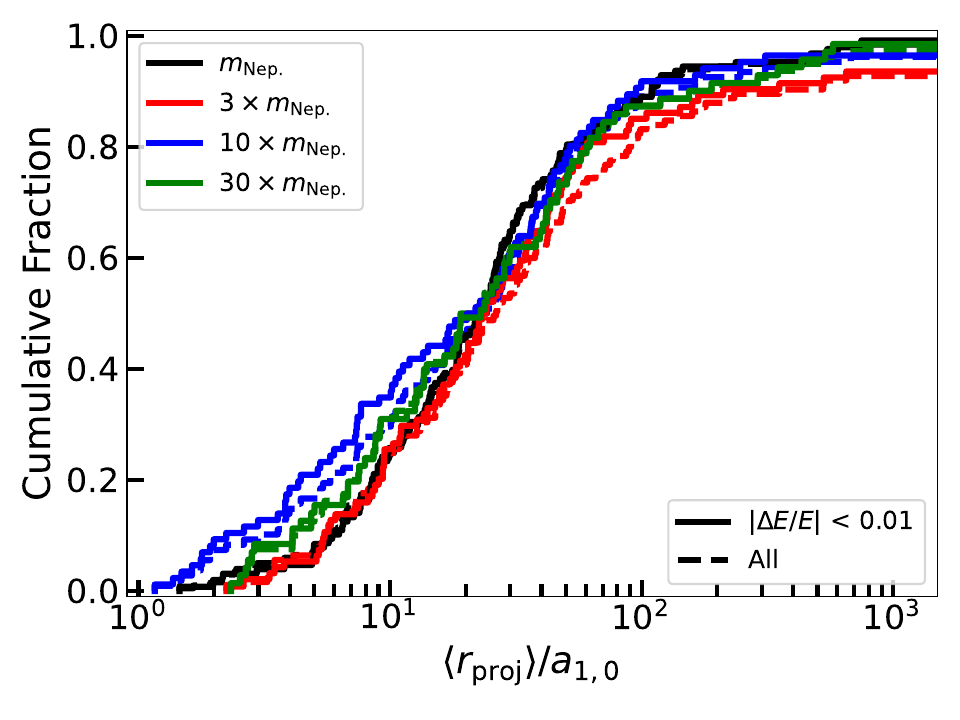}
    \caption{
    Cumulative distributions for the mean projected distances, $\langle r_\mathrm{proj} \rangle = \frac{\pi}{4}a\paren{1  + \frac{1}{2}e^2}$, between the detached planets and their host stars. Results for all simulations and only those with energy errors $|\Delta E/E|<0.01$ are shown with dashed and solid lines, respectively. Such planets can masquerade as 'free-floating' planets in microlensing surveys, when $a_{1,0}$ is beyond a few AU.}
    \label{fig:rproj}
\end{figure}

We now turn to consider the  projected radial separations of detached planets from their host stars, in anticipation of microlensing measurements. For a Keplerian orbit, the time-averaged radial distance is calculated as 
\begin{equation}
    \langle r \rangle = \frac{1}{2\pi}\int_{0}^{2\pi}a(1-e\cos u)\frac{dM}{du}du = a\paren{1+\frac{e^2}{2}}~.
\end{equation}
The average projected separation, assuming isotropic orientations of systems, is then $\frac{\pi}{4}\langle r \rangle = \frac{\pi}{4}a\paren{1+\frac{e^2}{2}}$.  Fig. \ref{fig:rproj} shows that  the detached planets are distributed roughly uniformly in log distance between $10a_{1,0}$ and $100a_{1,0}$, for all ensembles.
For $a_{1,0}\gtrsim$ a few AU or greater, these planets would be so far from the host stars, they would likely be identified as free-floating planets in microlensing surveys \citep{HanGaudi2005}. 

\begin{figure}
    \centering
    \includegraphics[width=0.9\columnwidth]{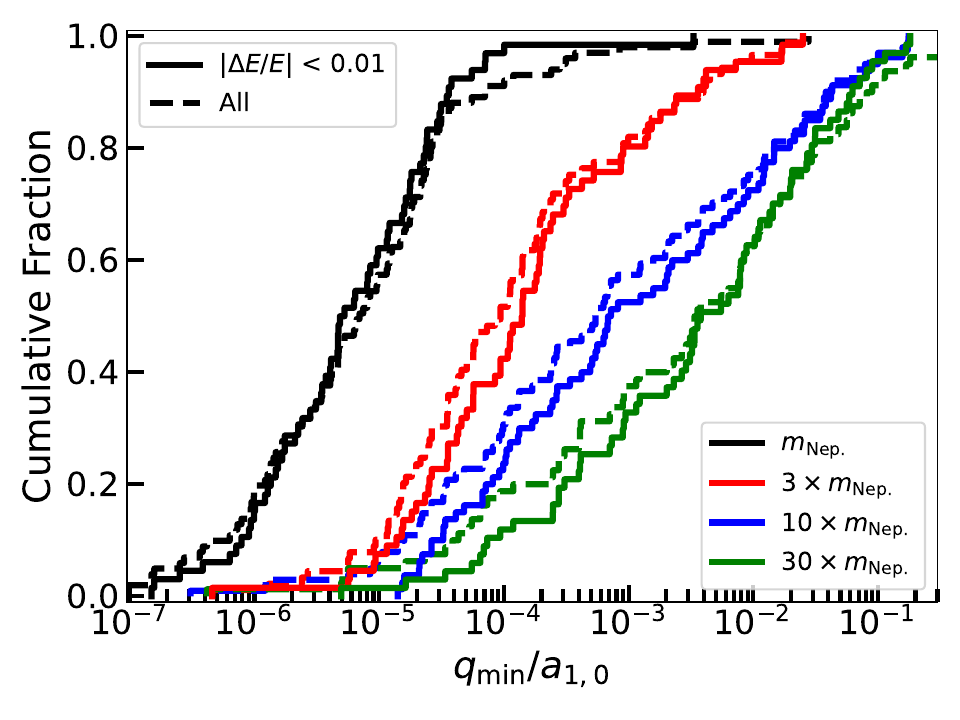}
    \caption{Cumulative histogram of the minimum pericenter distances reached by the innermost planets, measured in units of the initial semi-major axes of the innermost planets. Results for all simulations and only those with energy errors $|\Delta E/E|<0.01$ are shown with dashed and solid lines, respectively. Many could have collided with the central stars -- for $a_{1,0} = 10$AU, the stellar surface is at $\sim 5\times 10^{-4} a_{1,0}$. Ensembles of lower-mass planets can reach even smaller minimum distances.}
    \label{fig:qmin}
\end{figure}

Lastly, we investigate the orbits of the innermost planets, in particular, their minimum approaches to the host stars over the course of the integrations (Fig. \ref{fig:qmin}).
 It appears that these planets can encroach extremely close to their stars, with the lower mass ones reaching the lowest pericenter distances. The majority of the low-mass systems have closest approaches beneath a solar radius, for $a_{1,0} \leq 10$AU. We discuss the potential influence  this has on the fate of the inner systems in Section \ref{sec:discussion}.

Why are low-mass planets particularly susceptible to close pericenter passages? As we argue above, unstable systems eventually reach a dynamical regime where detached planets' orbital evolution is largely driven by secular angular momentum exchange, punctuated occasionally by close encounters with the inner planet. The timescale of secular angular momentum exchange scales inversely with planetary mass while the time-scale to fully eject detached planets scales more steeply with planet mass (e.g., Eq. \ref{eq:Teject}). 
Thus, planets in lower-mass systems live for a larger number of secular timescales. During this time, the pericenter of the innermost planet can reach especially small values through chaotic AMD exchange with the other planets.

\subsection{One Jupiter and ten Neptunes}
\label{sec:one_jupiter}
\begin{figure}
    \centering    
    \includegraphics[width=0.95\columnwidth]{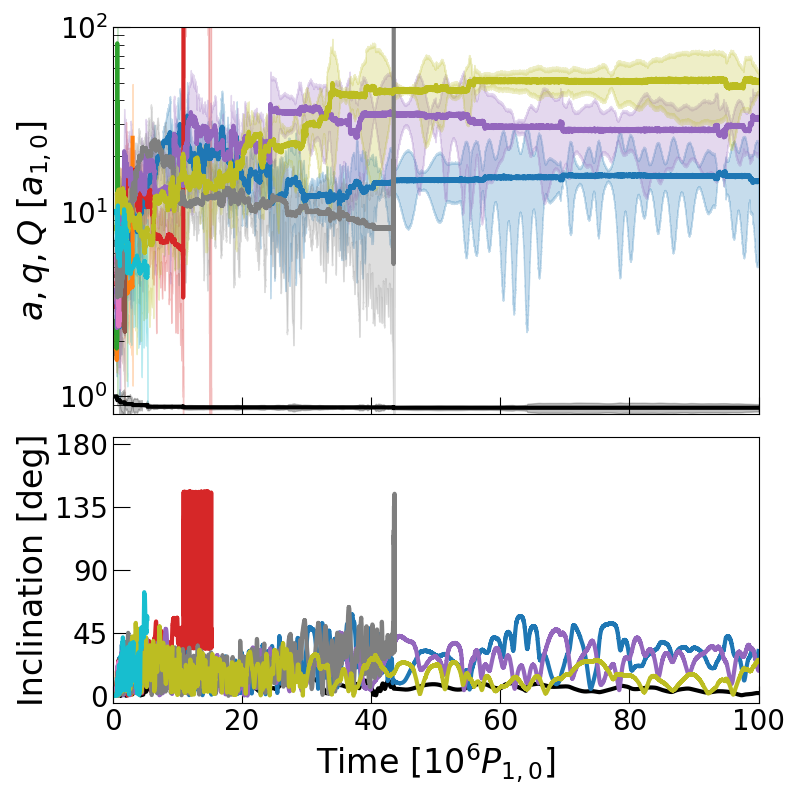}
    \caption{Similar to Figure \ref{fig:example} but for a system of one Jupiter-mass planet scattering 10 Neptunes (1J+10N). The orbit of the Jupiter-mass planet is shown in black.  
    Eventually, three Neptunes survive the onslaught and reach dynamically detached orbits.
    }
    \label{fig:jup_and_nep_example}
\end{figure}
Figure \ref{fig:jup_and_nep_example} shows the evolution of a typical  member of the 1J+10N ensemble. Before a few million orbits have elapsed, multiple Neptunes have been ejected by the Jupiter. Beyond this, however, the system's evolution dramatically slows down.  The remaining planets torque each other such that they lift up each other's pericenters and avoid close encounters with the Jupiter. A similar dynamical behaviour was observed in the equal-mass ensembles and has also been reported by \citet{silsbee_producing_2018}, who simulated sub-Earth mass embryos scattered by the solar system's giant planets. The remaining Neptunes become decoupled from the Jupiter-mass planet until chaotic quasi-secular exchanges of angular momentum push one of the planets into the path of the Jupiter, after which it is promptly ejected. This process removes AMD from the system until the remaining planets have too little AMD to allow further encounters. This occurs after $\sim50$ million orbits and when there are $3$ detached Neptunes.  These latter planets continue to interact, as evidenced by their modulating percienter distances. However, the perturbations are weak and the system's AMD is crudely conserved. Eventually, some of these planets would eject each other. But this would occur over too long a timescale to be of interest.

\begin{figure*}
\centering
\includegraphics[width=\linewidth]{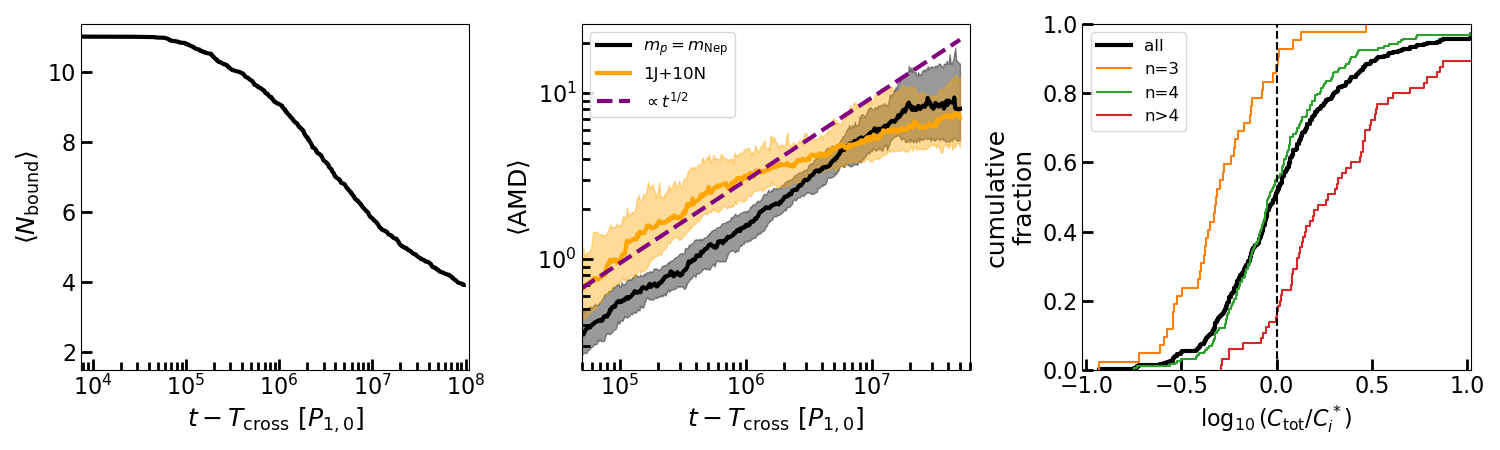}
\caption{
Similar to Figs. \ref{fig:bound_vs_t}-\ref{fig:amd_cdfs} but for the 1J+10N ensemble.
Left panel shows that, on average, the Jupiter is able to eject all but three Neptunes. 
In the middle panel, the average AMD for the 1J+10N systems (yellow) are compared against those in the 5 equal-mass ensemble (black). 
The behaviour is similar in both cases, suggesting that the concept of AMD stability controls the overall dynamics. 
Right panel:  colored lines show cumulative distributions of $C_\mathrm{tot}/C^*_i$
   for the subsets of systems that
   with $n=3$ (yellow), or $4$ (green), and $>4$ (red) 
   planets surviving, in the 1J+10N ensembles. On average, two detached planets survive.
  }
  \label{fig:combined}
\end{figure*}

Further details on the 1J+10N ensemble are provided in Fig. \ref{fig:combined}, where we exhibit the mean number of bound planets as a function of time, as well as the evolution of mean AMD.  Lastly, we also investigate the AMD stability of the detached planets. We find that  approximately half of the surviving detached planets are AMD-stable ($C_i^*/C_\mathrm{tot} < 1$)  at the end of $10^8$ orbits, while the rest are marginally unstable. Thus, we establish again that AMD stability is instrumental in determining the final state of the systems. 

However, the average number of detached planets in the 1J+10N ensemble is about one more than is found in the equal-mass ensembles. In particular, the final stable state for the former contains 2 detached planets, while only one for the latter. Why is this so?  We do not have the full answer at the moment.

\section{Discussion}
\label{sec:discussion}

\subsection{Can Known Planets Eject FFPs?}

Microlensing reports a rate of two free Neptunes per star. Can they be produced by planet-planet ejections? For our discussion here, we consider  M-type hosts only.

First of all, planets responsible for the production of FFPs must have sufficiently large Safronov numbers.
From Equation \eqref{eq:safronov},  we obtain the critical semi-major axis, $a_\mathrm{ej}$, beyond which $\Theta>1$,
\begin{eqnarray}
    a_\mathrm{ej} &= 2.4~\mathrm{AU}\fracbrac{q_\mathrm{Nep}}{q}^{2/3}\fracbrac{M_*}{0.5M_\odot}^{1/3}\fracbrac{2\mathrm{g/cm}^3}{\rho_p}^{1/3}~.
    \label{eq:safranov2}
\end{eqnarray}
Interestingly, the occurrence rates of planets in this orbital range have been characterized by RV, transit, and, most importantly, microlensing (see \S \ref{sec:intro}). This allows us to evaluate the scattering hypothesis quantitatively.

First, the Jovian planets.
While  they can easily eject, their low occurrence rate around M-dwarfs \citep[$5-6\%$,][]{zang_microlensing_2025,montet_trends_2014} makes them unlikely culprits. Each Jovian would have to scatter an unbelievably large number of  Neptunes  ($\sim 40$), in order to produce the $2$ free-floating Neptunes per star.  

Meanwhile, Neptune-class planets are about ten times more populous \citep[$\sim 0.35$ per star for the mass range $-4.5 < \log q < -3.5$;\footnote{This corresponds roughly to the same mass range as quoted for FFPs, when adopting a stellar mass of $0.5 M_\odot$. }][]{zang_microlensing_2025}.
If Neptune-class planets are the scatterers, one can explain
the occurrence rates of both bound and free-floating Neptune-sized planets  by supposing that one in every three stars forms with $\sim 7$ Neptune-mass planets within tens of AU. These systems would also often undergo dynamical instabilities. When this happens, we find that typically onw planet becomes more tightly bound (the `bound' planet), while the remainder are ejected or launched into very wide orbits (`detached'). These latter, bound or not, may appear as FFPs in microlensing studies. 

Four concerns with this scenario arise.
First, Neptune planets orbiting at $\sim 10$ AU may also masquerade as FFPs, as opposed to being the truly free or the detached ones (hundreds of AU) studied here. The sensitivity of microlensing surveys to the simultaneous detection of a host star and bound planet is sharply peaked at orbital separations near the Einstein radius, 
\begin{multline}
    R_E = 3\mathrm{AU}~\fracbrac{0.5M_\odot}{M_*}^{1/2}\fracbrac{8~\mathrm{kpc}}{D_\mathrm{source}}^{1/2}\fracbrac{x(1-x)}{0.25}^{1/2},
    \label{eq:Re}
\end{multline}
 where $x=D_\mathrm{lens}/D_\mathrm{source}$.
So the stellar lensing signature may be missing for bound planets orbiting beyond a few $R_E$, and they can be mistaken as FFPs. 
But to explain the $2$ Neptunes per star number, the number of planets in this range would have to be $\sim 10$ times above those within a few AU. This may be hard to arrange if we adopt an outer radius for planetary systems to be $\sim40$AU, typical of protoplanetary disk sizes.

Second, many of the bound planets may turn out to be in multi-planet systems. Today, there are a handful of two-planet lenses discovered among the two-hundred strong sample \citep[see, e.g.,][]{Gaudi2008}, and (likely due to selection effects) they are almost exclusively Jovian pairs \citep[see Fig. 8 of][]{Li2025}. This frequency is surprisingly  high considering the strong bias against detecting multiple planets in microlensing, and suggests that multiplicity is common. If such a conclusion also applies to Neptune-massed planets, it makes the ejection hypothesis less tenable, since it is hard to retain two or more closely-spaced planets after a full-scale dynamical instability.

Our third and fourth concerns relate to the mass functions of FFPs.
Lower-mass FFPs (below $7 M_\oplus$) are observed to be even more abundant than Neptunes and can carry comparable masses \citep{sumi_free-floating_2023}. They are unlikely to be ejected by planets of similarly low masses, both because of the exceedingly long timescale involved (Eq. \ref{eq:Teject}) and because of their lower Safranov numbers (Eq. \ref{eq:safronov}).
They could be ejected by higher mass planets like Neptunes.
However, such a scenario poses a cosmogenic tension: since only about one in three systems contains a bound Neptune within a few AU \citep{zang_microlensing_2025}, it would require these systems to produce almost all the planets (bound or otherwise), while the other two are barren. This tension may be relieved if at least some of these lower mass FFPs are instead orbiting at tens of AUs.

Lastly, one may ask whether the observed mass functions of bound and free planets support the ejection scenario. In a given system, the most massive planet is likely to stay bound. So the FFPs should favour lower masses than the bound ones.  This may account for  the claimed differences in  the mass functions for the  FFPs (Eq. \ref{eq:sumi}) and  for the bound planets (Eq. \ref{eq:zang}). However, the data is still too sparse to be certain.
 
\subsection{Many FFPs are merely `detached'}

An interesting result from our simulations is that scattering dynamics not only ejects planets but also produces `detached' objects that orbit at tens to hundreds of times their original distances. Detached objects arise because secular torques among the scattered planets extricate them from encountering the scatterer. While later AMD exchanges can still remove some of these planets (Fig. \ref{fig:example}), we find frequently that a couple planets remain `detached'  but bound. At the end of our integrations, the average number of detached planets per system is $\sim 2$ for our Neptune-mass ensemble and $\sim 3$ detached Neptunes for the 1J+10N simulations. 

We can estimate an occurrence rate for `detached' Neptunes by assuming that all known planetary systems are mobilized in scattering.
With the occurrence rates of $\sim 0.35$ bound Neptunes per star and $\sim 0.06$ Jovians per star, the number of detached Neptunes per star is
\begin{equation}
N_{\rm detached} \approx 2 \times 0.35 + 3 \times 0.06 \approx 0.9 \, ,
\end{equation}
or about half of the reported Neptune-mass FFPs. This rate could reduce to $\sim 0.7$ Neptunes per star if the Jovian systems are not engaged in scattering, and reduce further to $\sim 0.2$  in the unlikely case that only the Jovian systems are responsible for ejections. This rate can increase, on the other hand, if a fraction of the FFPs are simply planets orbiting at tens of AU, not results of scattering. They are bound to the host stars and so can also be categorized as `detached'.

As Fig. \ref{fig:rproj} shows,  detached bodies orbit between $10-100$ times the original separation  (median at $\sim 30$ times). If the original $a_{1,0} = 10$AU, we expect the new separation to be $\sim 300$AU, or an angular separation of $\sim 0.07"$ if the system is at 4~kpc. This lies well outside the Einstein radius of a star (Eq. \ref{eq:Re}) and these planets would most likely be classified as FFPs when they happen to lens. 

Confirming the detached nature of some of the FFPs would then provide evidence for the ejection scenario. There may be two ways to do so. One is  to look for rare events 
 where the trajectory of a background source intersects the Einstein radii of both the host star and the detached planet. For a typical lens at 4~kpc, a source at 8~kpc, a $10~M_\oplus$ planet would have an Einstein radius of $\sim 6~\mu\mathrm{as}$, or $\sim 10^{-4}$ times the planet-star separation if it orbits at 300~AU around the host. We then predict an event rate of $\sim 0.35 \times 10^{-4}/\pi \sim 10^{-5}$ per stellar lensing event. This is difficult as it requires two orders of magnitude more events than in our current collection. 

More probably, one can use the fact that a host star should lie at $\sim 0.1"$ away from the detached planet when the latter acts as a lens. A high  contrast image could detect the star near the time of the planet lensing. Or one can look for the stellar signature well preceding or succeeding the planet lensing event, as was recently attempted by \citet{Kapusta2025}. Unfortunately, 
their attempt was not successful.

The number of detached bodies can also be affected by  galactic tides and flyby stars. Analogous to the emplacement of scattered planetesimals into the Oort cloud \citep{Duncan1987}, slow secular torques from the galactic tide could lift the pericenters of distant planets, decoupling them from the scatterings and stranding them in the distant outskirts of the planetary system \citep[see, e.g.][]{raymond_oort_2023}. Fly-by stars can provide similar torques, as well as impulsive kicks that ionize weakly-bound planets from the system {\citep[see, e.g.,][]{boley_interactions_2012}}.

\subsection{Effects of and Consequences for Inner Systems}

While planetary systems may extend multiple decades in orbital separation, we have  only examined one decade in isolation.
Transit surveys have revealed that most, if not all, early M-dwarf stars host close-in planets  inwards of an AU \citep[see, e.g.][]{hsu2020}. 
While these planets may not be massive enough to eject each other (Eq. \ref{eq:safranov2}), they will nevertheless be impacted when  planets much further out  are driven to low pericenter distances by scattering. 
Figure \ref{fig:qmin} shows that, if we adopt  $a_{1,0} =  10~\mathrm{AU}$ then the inner-most planets in nearly all simulated systems  can reach $0.1$ AU or closer, with 
those in the  $m_p= m_\mathrm{Nep}$ and $3~m_\mathrm{Nep}$ ensembles  approaching the star to within one stellar radius. If these planets are either engulfed or tidally captured by their central stars, further ejection of any remaining planets will be slowed down and more detached planets may result.

However, none of these possible effects are captured by our idealized simulations. We have also not included effects like scattering/colliding with the inner planets, orbital precession by general relativity and rotational/tidal bulges. Therefore, we are at no position to predict the actual dynamics.  But it does raise the possibility of obtaining further evidence for dynamical upheavals in the outer regions of planetary systems by examining inner systems.

We add in one speculative remark regarding the so-called `super-Earth/cold-Jupiter' correlation
\citep[e.g.,][]{ZhuWu2018,bryan_excess_2019}   where inner super-Earths and outer cold Jupiters appear to occur in tandem. In particular, \citet{ZhuWu2018} inferred a conditional occurrence rate of $90\%$ for super-Earths  in systems that host cold Jupiters. Perhaps these cold Jupiters, being relatively immobile  and highly efficient at ejection, protect the inner systems from being dynamically disrupted by interlopers coming from the unstable outer systems.  

\section{Conclusions}
We have explored the dynamics of planet-planet scattering and its consequences for the production of free-floating and wide-orbit planets using  ensembles of $N$-body simulations. Our main conclusions are as follows:
\begin{enumerate}
    \item The timescale over which planets are ejected increases steeply with decreasing planet mass (Eq. \ref{eq:Teject}). In particular, Neptune-mass planets at a few
     AU and beyond can require billions of years to dynamically  evacuate.  In fact, the dynamical havoc is still on-going for even ancient systems.
    \item Scattering does not remove all planets.  We find that
     a single tightly bound planet and 1-3 ``detached" planets at large orbital distances ($\sim10–100\times$ the initial separations)  are left at the end of our simulations.  These detached planets would almost always appear as FFPs in microlensing observations. We suggest that about half of the reported Neptune-mass FFPs should be instead merely `detached'. This is testable.
    \item  Angular momentum deficit (AMD) plays a key role in dictating systems' late-time dynamical evolution and setting the ultimate orbital configurations. 
    The systems reach their final states (one inner planet plus one detached planet) when there is no longer sufficient AMD to cause orbit crossing.    
    A deeper theoretical understanding of the AMD evolution is needed.    
    \item   The inner planetary system (inwards of a few AU) may also suffer from the consequences of the scatterings further out. The smallest peri-centres of the scattering planets can dip well below an AU. This likely has dire, and observable, consequences. 
    
\end{enumerate}
Overall, our results suggest that the ejection scenario for FFPs naturally predicts a residual population of wide-orbit planets, and that outer system dynamics may leave detectable imprints on both microlensing statistics and the architectures of inner planetary systems.

\begin{acknowledgements}
S.H. acknowledges support by the Natural Sciences and Engineering Research Council of Canada (NSERC), funding references CITA 490888-16 and RGPIN-2020-03885; YW acknowledges NSERC grant RGPIN-2024-05533 and thanks Shude Mao, Weicheng Zang for discussions.
\end{acknowledgements}

\bibliography{references}
\bibliographystyle{aasjournal}

\end{document}